\documentclass[aps,prx,twocolumn,superscriptaddress,citeautoscript,amssymb,reprint,showpacs,floatfix]{revtex4-1}
\usepackage[ansinew]{inputenc}
\usepackage[T1]{fontenc}
\usepackage{float}
\usepackage{amsmath}
\usepackage{ulem}
\usepackage{mathptmx}
\usepackage{mathrsfs}
\usepackage[english]{babel}
\usepackage{graphicx}
\usepackage{color}
\usepackage{natbib}

\begin{document}
\author{J. F. Landaeta}
\thanks{correspondence should be addressed to javier.landaeta@cpfs.mpg.de or elena.hassinger@cpfs.mpg.de}
\affiliation{Max Planck Institute for Chemical Physics of Solids, 01187 Dresden, Germany}

\author{P. Khanenko}
\affiliation{Max Planck Institute for Chemical Physics of Solids, 01187 Dresden, Germany}

\author{D.C. Cavanagh}
\affiliation{Department of Physics, University of Otago, P.O. Box 56, Dunedin 9054, New Zealand}

\author{C. Geibel}
\affiliation{Max Planck Institute for Chemical Physics of Solids, 01187 Dresden, Germany}

\author{S. Khim}
\affiliation{Max Planck Institute for Chemical Physics of Solids, 01187 Dresden, Germany}

\author{S. Mishra}
\affiliation{Laboratoire National des Champs Magn\'etiques Intenses (LNCMI-EMFL), CNRS, UGA, 38042 Grenoble, France}	

\author{I. Sheikin}
\affiliation{Laboratoire National des Champs Magn\'etiques Intenses (LNCMI-EMFL), CNRS, UGA, 38042 Grenoble, France}	

\author{P.M.R. Brydon}
\affiliation{Department of Physics and MacDiarmid Institute for Advanced Materials and Nanotechnology, University of Otago, P.O. Box 56, Dunedin 9054, New Zealand}

\author{D.F. Agterberg}
\affiliation{Department of Physics, University of Wisconsin-Milwaukee, Milwaukee, Wisconsin 53201, USA}

\author{M. Brando}
\affiliation{Max Planck Institute for Chemical Physics of Solids, 01187 Dresden, Germany}

\author{E. Hassinger}
\thanks{correspondence should be addressed to javier.landaeta@cpfs.mpg.de or elena.hassinger@cpfs.mpg.de}
\affiliation{Max Planck Institute for Chemical Physics of Solids, 01187 Dresden, Germany}
\affiliation{Technical University Munich, Physics department, 85748 Garching, Germany}

\title{Field-angle dependence reveals odd-parity superconductivity in CeRh$_2$As$_2$}
\date{\today}

\begin{abstract}
CeRh$_2$As$_2$ is an unconventional superconductor with multiple superconducting phases and $T_\mathrm{c} = 0.26$\,K. When $H\parallel c$, it shows a field-induced transition at $\mu_0H^* = 4$\,T from a low-field superconducting state SC1 to a high-field state SC2 with a large critical field of $\mu_0H_\mathrm{c2} = 14$\,T.  In contrast, for $H\perp c$, only the SC1 with $\mu_0H_\mathrm{c2} = 2$\,T is observed. A simple model based on the crystal symmetry was able to reproduce the phase-diagrams and their anisotropy, identifying SC1 and SC2 with even and odd parity superconducting states, respectively. However, additional orders were observed in the normal state which might have an influence on the change of the superconducting state at $H^*$.
Here, we present a comprehensive study of the angle dependence of the upper critical fields using magnetic ac-susceptibility, specific heat and torque on single crystals of CeRh$_2$As$_2$. The experiments show that the state SC2 is strongly suppressed when rotating the magnetic field away from the $c$ axis and it disappears for an angle of 35$^{\circ}$. This behavior agrees perfectly with our extended model of a pseudospin triplet state with $\vec{d}$ vector in the plane and hence allows to nail down that SC2 is indeed the suggested odd-parity state.
\end{abstract}
\maketitle
\section{Introduction}
Odd-parity, or (pseudo-) spin triplet superconductivity is a rare phenomenon in nature. Only a few materials are promising candidates, including UPt$_3$ \cite{Fisher1989}, the ferromagnetic superconductors UCoGe, URhGe, UGe$_2$\cite{Aoki2019}, and UTe$_2$ \cite{Ran2019}. In these systems, important information on the superconducting (SC) state has come from an investigation of the angle dependence of the critical fields. Theoretically, when spin-orbit coupling is strong, the angle dependence of the Pauli-limiting field $H_p$ offers a way to identify possible triplet states since $H_p$ should depend on the orientation of the applied magnetic field with respect to the direction of the spins of the Cooper pairs \cite{Sigrist2005,Machida1985,Mineev1999}. In experiment however, the observed angle dependencies of the upper critical fields in the above mentioned systems seem to be dominated by the orbital limit or by the interplay of superconductivity and field-enhanced spin fluctuations associated with an Ising quantum critical point \cite{Aoki2019,Tada2011}. 

Recently, CeRh$_2$As$_2$ was discovered to have unique and highly anisotropic superconducting critical field phase diagrams (Fig. \ref{fig:Figure_4}a,f), with a suggested field-induced odd-parity state \cite{Khim2020}. The aim of this study is to investigate its angle dependence in detail in order to find out more about the superconductivity itself and its relation with the normal state.

CeRh$_2$As$_2$ is a heavy-fermion compound with an electronic specific heat coefficient close to 1\,J/molK$^2$ at 0.5\,K.  At $T_0 \approx~0.4$\,K, a phase transition to a suggested quadrupole-density-wave (QDW) state occurs~\cite{Hafner2021}. At $T_\mathrm{c}=0.26$\,K, CeRh$_2$As$_2$ enters a low-field superconducting state (SC1). A magnetic field applied along the $c$ axis induces a transition at $\mu_0H^*\approx 4$\,T into a second high-field superconducting state SC2, with an upper critical field $H_{c2} = 14$\,T. For in-plane fields, only the low-field state SC1 appears, with $H_{c2} = 2$\,T, whereas $T_0$ increases with field so that the SC1 phase is found to be always included in the QDW phase for this field direction. Furthermore, in zero field within the SC state a broadening  of  the  As(2) nuclear quadrupole resonance (NQR)  line   was  interpreted as the onset of antiferromagnetic (AFM) order \cite{Kibune2021}. However, no additional anomaly was detected in the bulk measurements such as specific heat and thermal expansion \cite{Khim2020,Hafner2021}.

The crystal structure of CeRh$_2$As$_2$ is centrosymmetric, but the inversion symmetry is broken locally at the Ce sites enabling a Rashba spin orbit coupling with alternating sign on neighboring Ce layers \cite{Khim2020}. Assuming dominant intralayer singlet SC pairing, a $c$-axis field transforms an even-parity superconducting state with equal gap sign on both Ce layers into an odd-parity one with alternating gap sign \cite{Yoshida2012,Khim2020}. The latter may be topological \cite{Skurativska2021,Nogaki2021} and can be described as a pseudospin triplet state with $\vec{d}$ vector in the plane leading to the absence of a Pauli limit for $H\parallel c$ \cite{Khim2020, Yoshida2014}. 

The unusual results for the field along the $c$ axis may also have another origin. One possibility emerged from the observation that the QDW state below $T_0 \approx 0.4$\,K is suppressed by a $c$-axis field of similar strength as $H^*$ \cite{Khanenko2022,Khim2020,Hafner2021}. So, the SC1 phase is found to be inside the QDW phase as for in-plane fields, but the SC2 phase would be outside the QDW phase. The presence of the QDW state might lead to a drastic change of the superconducting properties and in particular of $H_{c2}$ between the SC1 and SC2 states~\cite{Khim2020}. Ignoring the QDW and AFM phases, other possibilities have been suggested to explain the origin of the SC1 and SC2 phases as a change between different superconducting order parameters ~\cite{Skurativska2021, Moeckli2021, Moeckli2021-2}. There are scenarios which involve a field driven Fermi-surface Lifshitz transition \cite{Ptok2021} or the presence of a field-induced quantum critical point (QCP)~\cite{Aoki2019,Tada2011}. Eventually, because the normal state is also anisotropic, a study of the angle dependence of the superconducting states is a promising way to distinguish between those scenarios. 

Hence, we study the superconducting phase diagram as a function of magnetic field direction by ac susceptibility, magnetic torque and specific heat. We find that the angle dependence supports the picture that SC2 is an odd parity state with pseudo-spin $\vec{d}$ vector in the plane. CeRh$_2$As$_2$ seems to be the first compound where the anisotropy of the Pauli limiting can be used to reveal triplet superconductivity.

\section{Results}
In the following, we show results of ac susceptibility $\chi_{ac}$, magnetic torque $\tau$ and specific heat $C$ as a function of temperature and magnetic field in different directions with an angle $\theta$ away from the $c$ axis. These probes were chosen because they are sensitive to the bulk properties of the material and they can detect the transition inside the superconducting state \cite{Khim2020}. Previously it was found that the $T_\mathrm{c}$ from resistivity is higher than the bulk $T_\mathrm{c}$ from specific heat or low-frequency ac-susceptibility \cite{Khim2020}. It was suggested that this is due to percolating strain-induced superconductivity around impurities. Experimentally, the resistive $T_\mathrm{c}$ follows the bulk $T_\mathrm{c}$ in a parallel fashion for both $H\parallel c$ and $H\parallel a$ \cite{Hafner2021,Onishi2022}. While the angle dependence of the superconducting state could have been investigated by resistivity as well, the transition at $H^*$ is invisible to this probe \cite{Khim2020,Hafner2021,Onishi2022}.

\begin{figure}[t]
	\centering
	\includegraphics[width=\linewidth]{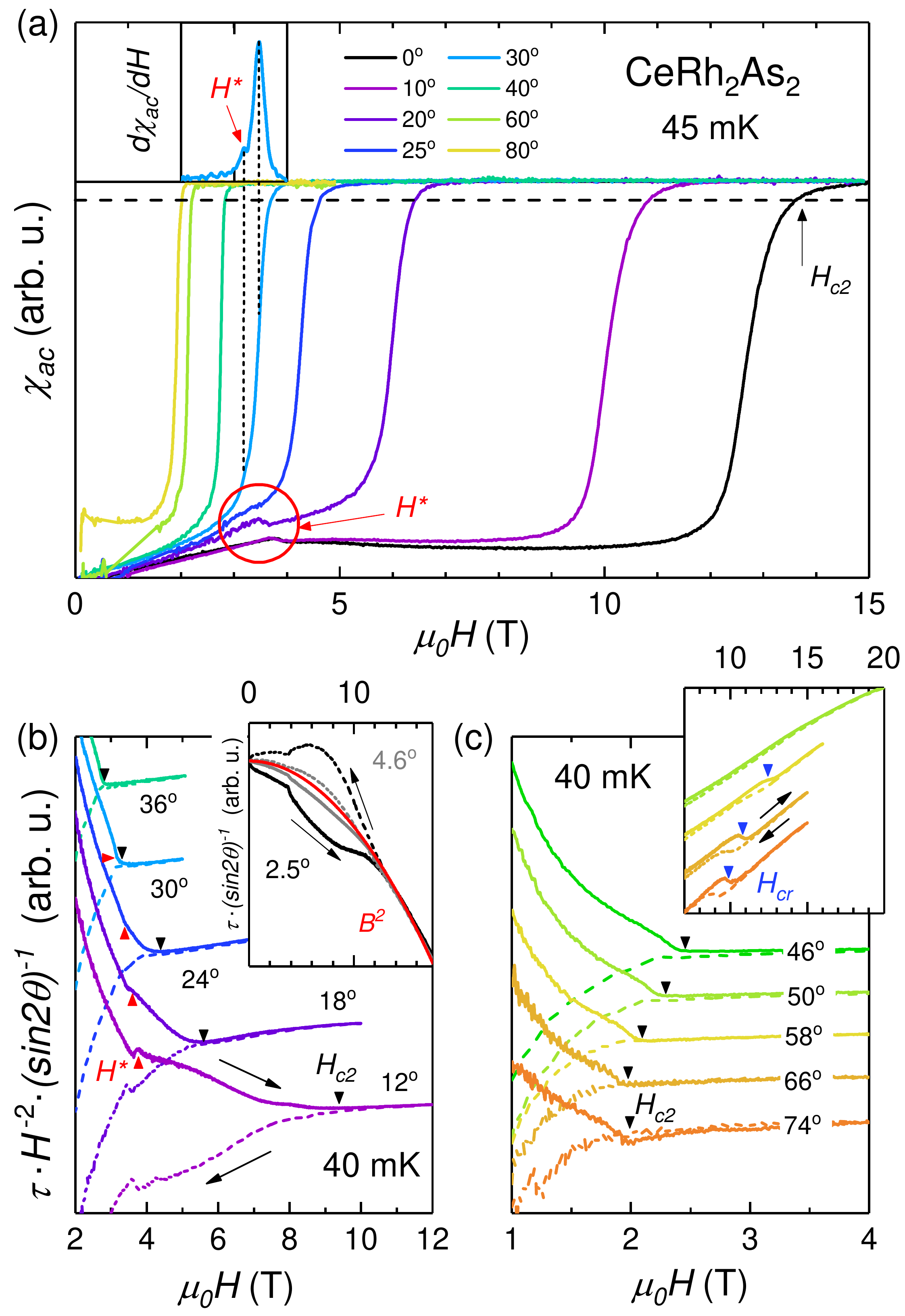}
	\caption{{\bf Magnetic field dependence of magnetic susceptibility and torque for different field directions.} (a)  Magnitude of the ac susceptibility $\chi_{ac}$ at 45\,mK for different angles as indicated. $\theta=0^\circ$ corresponds to $\mu_0 H\parallel c$. The bump labelled as $H^*$, is the transition between SC1 and SC2. The inset shows the derivative of the curve at 30$^\circ$, indicating that $H^*$ is still present at that angle. The dashed horizontal line indicates the value at which $H_\mathrm{c2}$ is defined. (b) Magnetic torque divided by $H^2\sin(2\theta)$ at 40\,mK. The inset shows the torque for $2.5^\circ$ and $4.6^\circ$ where the change in the hysteresis of $\tau$ is visible. Additionally, the red line gives the expected $B^2$-dependence. (c) Torque divided by $H^2\sin(2\theta)$ for angles close to the plane. The inset in (c) shows the additional phase transition appearing at $H_\mathrm{cr}\approx 9$\,T. In (b) and (c) curves are shifted for clarity.}
\label{fig:Figure_1}
\end{figure}

Figure \ref{fig:Figure_1}(a) shows the magnetic field dependence of $\chi_{ac}$ at the temperature of 45\,mK for different angles. Details on the experimental methods can be found in \cite{SM2021}. We measured from $\theta=0^\circ$ ($H\parallel c$) to $90^\circ$ ($H\perp c$ here named $H\parallel ab$). The in-plane field orientation was [110] for torque but not defined for $\chi_{ac}$ and $C/T$. A small but clear signature of the transition between SC1 and SC2 named $H^*$ appears for angles below 35$^\circ$. At 30$^\circ$ a small anomaly is visible at $H^*$ in the derivative $d\chi_{ac}/dH$ highlighted in the inset of Figure \ref{fig:Figure_1}(a).

We define the upper critical field $H_\mathrm{c2}$ at the onset of the diamagnetic transition in order to be consistent with the previous analysis \cite{Khim2020} (horizontal dashed line in Figure \ref{fig:Figure_1}(a)) and $H^*$ at the maximum of the small bump in the field dependence of $\chi_{ac}$.
The torque $\tau$ is shown in Figure \ref{fig:Figure_1}(b,c). As expected, its field dependence depicted in the inset of panel (b) is quadratic in field in the normal state. Since we are interested in deviations from this behavior in the superconducting state, we present data as $\tau/H^2$ and scaled by $\sin{(2\theta)}^{-1}$ which is the standard angle dependence of torque. It displays a similar step-like feature at $H^*$. The strong hysteresis loop in the superconducting state was already observed in the magnetization in CeRh$_2$As$_2$ \cite{Khim2020,SM2021}. While the hysteresis is counter-clockwise at small angles (Inset of Fig. \ref{fig:Figure_1}b), as in the magnetization, it changes to a clockwise one at higher angles (Fig. \ref{fig:Figure_1}b). This is related to a change in the magnetization anisotropy (see \cite{SM2021}). $H_\mathrm{c2}$ is defined in the torque at the field where up and down sweep curves start to separate. Since this happens rather smoothly for small angles, there is a larger uncertainty on these points. Strictly speaking, this is the so-called "irreversibility field". As usually observed, it is slightly lower than $H_\mathrm{c2}$ from $\chi_{ac}$, but it follows its angle dependence.

For fields near the in-plane direction, an additional transition is observed at $H_\mathrm{cr} \approx $ 9\,T (Fig. \ref{fig:Figure_1}c and \cite{SM2021}), as observed before \cite{Hafner2021}, where it was associated with a change of the order below $T_0$. $H_\mathrm{cr}$ increases when the field is turned away from the $ab$ plane towards the $c$ axis until, for angles below 58$^\circ$, this anomaly cannot be resolved experimentally any more. As can be seen in the inset, it is strongly hysteretic indicating a first order transition, in agreement with results from resistivity and magnetostriction \cite{Hafner2021}. No further transitions have been detected for in-plane fields up to 36\,T and $c$-axis fields up to 26\,T \cite{SM2021}.

\begin{figure*}[t]
	\centering
	\includegraphics[width=\linewidth]{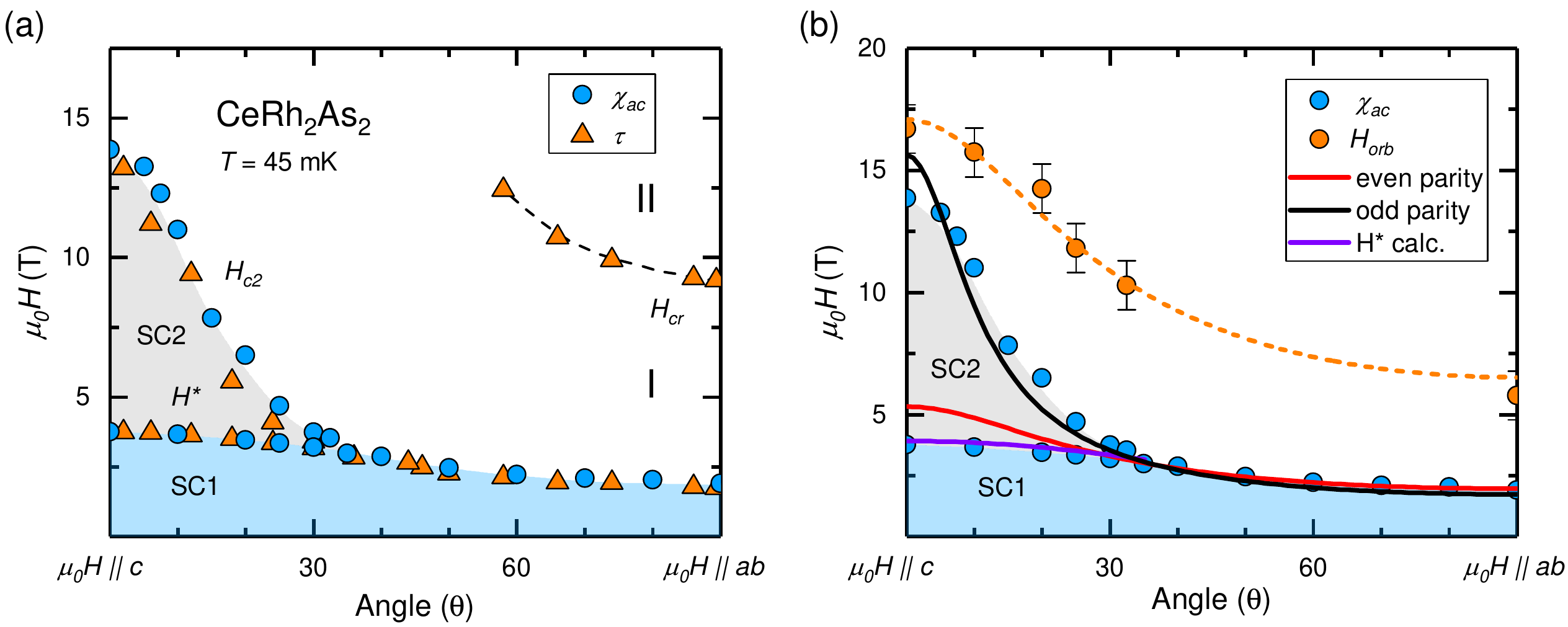}
	\caption{{\bf Angle dependence of the critical fields at 45\,mK}. (a) different symbols represent different experimental techniques as indicated. $H_\mathrm{cr}$ indicates a transition between two ordered states I and II \cite{Hafner2021}. Since it shows a strong hysteresis, only the critical fields from up sweeps are shown here. The critical fields from the down sweeps are typically around 1\,T lower. (b) Angle dependence of the upper critical fields of SC1 and SC2 as well as the orbital limit $H_\mathrm{orb}$ of SC1 obtained from the phase diagrams in Fig. \ref{fig:Figure_4} and a fit of it using equation \ref{eq:Hc_in_angle} (orange dashed line). Red and black lines are fits to equation \ref{eq:2} at 45\,mK and the violet line reflects the calculated $H^*$, as detailed in the main text and \cite{SM2021}.}
\label{fig:Figure_2}
\end{figure*}

From these data we find the angle dependence of $H_\mathrm{c2}$, $H^*$ and $H_\mathrm{cr}$ shown in Figure \ref{fig:Figure_2}(a). Here, the upper critical field $H_\mathrm{c2}$ decreases rapidly as the angle is increased with respect to the $c$ axis until it approaches the value of $H^*$ at about 35$^\circ$. From this point on, we only detect one superconducting state in which the $H_\mathrm{c2}$ slowly decreases with angle until it reaches the expected value of $\approx 2$\,T for $H\parallel ab$.

\begin{figure}[b]
	\centering
	\includegraphics[width=\linewidth]{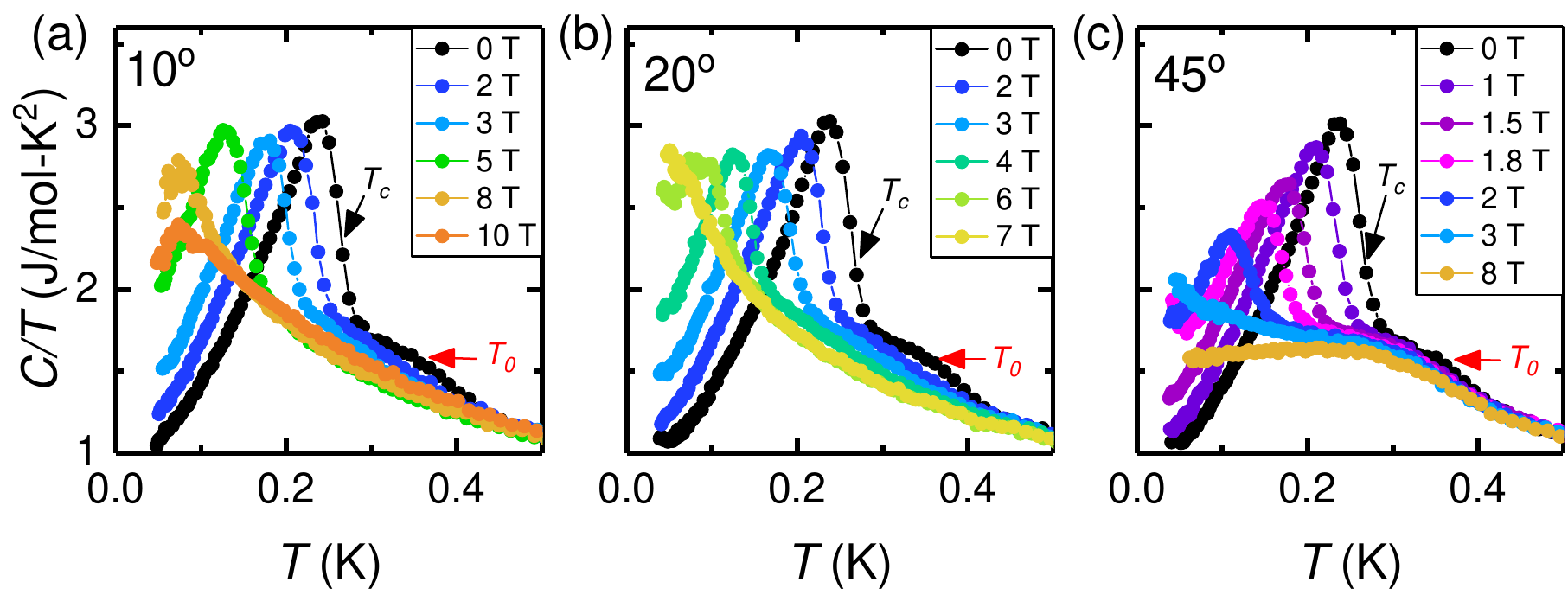}
	\caption{\textbf{Temperature dependence of the specific heat $C/T$ in magnetic field at different angles.} Here, a nuclear contribution has been removed \cite{SM2021}.}
\label{fig:Figure_3}
\end{figure}

For different magnetic field angles, temperature dependent ac-susceptibility (data in \cite{SM2021}) and specific heat were measured in magnetic field. A selection of the specific heat data are depicted in Fig. \ref{fig:Figure_3}, where a nuclear contribution has been removed. In this paper, these data are used to extract the superconducting transition temperatures and a qualitative behavior of $T_0$ (see below). A more detailed analysis of the specific heat and the angle dependence of $T_0$ is subject of future studies.

\begin{figure}[t]
	\centering
	\includegraphics[width=\linewidth]{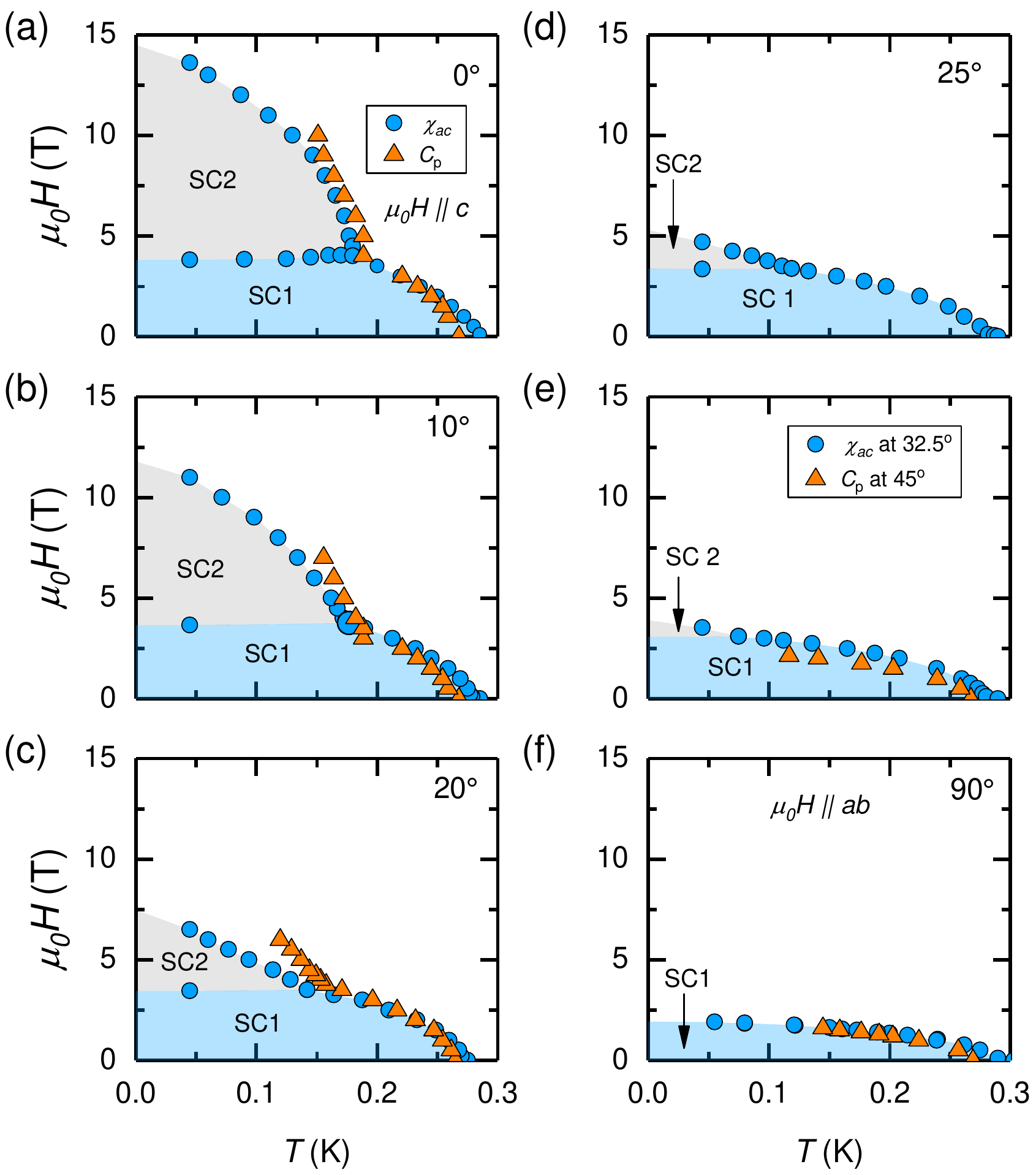}
	\caption{{\bf Magnetic field-temperature phase diagrams of CeRh$_2$As$_2$ for different directions of the magnetic field}. The experimental points are from ac susceptibility and specific heat, as indicated. The data show in (a) and (f) were taken from \cite{Khim2020}.}
\label{fig:Figure_4}
\end{figure}

The superconducting critical fields phase diagrams are shown in Figure \ref{fig:Figure_4} where data from the temperature sweeps in Fig. \ref{fig:Figure_3} and from \cite{SM2021} and from the field sweeps in Fig. \ref{fig:Figure_1} are included. Note that the discrepancy between the critical field from specific heat and ac-susceptibility at $20^\circ$ is ascribed to a difference in angle within the uncertainty because in this angle region, $H_\mathrm{c2}$ changes strongly even with only a few degrees. For all angles below 35$^\circ$, a kink appears in $H_\mathrm{c2}(T)$ at a field that is close to $H^*$ obtained from the field-sweeps revealing that the $H^*$ line is almost temperature independent for all angles. As already observed for the high-symmetry directions \cite{Khim2020}, the phase diagrams determined by ac-susceptibility are fully consistent with those obtained from the specific heat and hence reveal the bulk properties.

The present torque and specific heat data also provide some new information on the phase diagram of the phase connected with $T_0$  which was identified with a QDW order~\cite{Khim2020,Hafner2021}. An order of magnetic dipoles was ruled out since magnetic probes like ac-susceptibility or magnetization do not display any signature at this transition~\cite{Khim2020}. On the other hand, the small hump in the specific heat and a clear jump in the thermal expansion reveal a bulk phase transition. The electrical resistivity shows an increase below $T_0$ possibly due to nesting. This indicates that itinerant $f$-electrons forming the bands at the Fermi energy are involved in the $T_0$ order. The field dependence of $T_0$ is extremely anisotropic. For in-plane fields, an increase of $T_0$ is observed~\cite{Khim2020,Hafner2021}. Since such an increase is typical for local quadrupolar order in cubic systems, it was one of the signs that the $T_0$ order involves quadrupolar degrees of freedom of the itinerant electrons that are induced by a Kondo mixing of the crystal electric field doublets. This idea was supported by calculations of the quadrupole moment for such a system \cite{Hafner2021}.  Furthermore, at $\mu_0H_\mathrm{cr} = 9$\,T, a first-order transition to another state occurs. In contrast, for $c$-axis fields, the $T_0$ order is suppressed at roughly $H_0 = (4 \pm 1)$\,T, seemingly coincident with the transition inside the superconducting state \cite{Khim2020}. Both $H_0$ and $H_\mathrm{cr}$ present a clear angle dependence. For $10^{\circ}$ and $20^{\circ}$, the hump at $T_0$ is suppressed with field, but remains visible up to roughly 4\,T (Figure \ref{fig:Figure_3}a,b), similar to what was observed for $H \parallel c$ \cite{Khim2020}. For $45^{\circ}$ $T_0$ stays approximately constant in applied fields up to 10\,T (Figure \ref{fig:Figure_3}c), far above the superconducting critical field, similar to the behavior observed for $H\parallel ab$ \cite{Khim2020,Hafner2021}. Therefore, $H_0$ stays roughly constant at least up to $20^{\circ}$ and then starts increasing strongly for angles between $20^{\circ}$ and $45^{\circ}$. Furthermore, because $H_\mathrm{cr}$ shoots up for angles from $90^\circ$ to $58^\circ$, it might meet the $H_0$ line in a tricritical point in the range $30^\circ < \theta < 60^\circ$ and $\mu_0 H> 15$\,T. Another possibility is that the $H_\mathrm{cr}$ transition ends in a critical end point, since the anomaly just disappears going from $58^\circ$ to $50^\circ$. \\The absence of any further anomaly in the torque up to $\mu_0 H = 36$\,T for $\theta = 89.4^\circ$ ($H\parallel a$) \cite{SM2021} indicates that the phase forming at $H>H_\mathrm{cr}$ is stable until very high field as observed e.g. for the quadrupolar phase II in CeB$_6$ in high magnetic fields \cite{Effantin1985}. Further high-field studies are necessary to determine the $H_0$ boundary. \\Notably, the angle dependence of $H_0$ and $H_\mathrm{cr}$ are quite different from those of $H_\mathrm{c2}$, suggesting the interaction between this phase and superconductivity to be weak. On the other hand, for low angles, $H_0$ nearly coincides with $H^*$ up to $20^\circ$. \\


Let us now concentrate on the angle dependence of the critical field of the state SC1. Previous analysis along the two crystallographic directions showed that the orbital limit $H_\mathrm{orb}$ in CeRh$_2$As$_2$ is large and strongly exceeds the experimental critical fields of SC1. Therefore, this state is strongly Pauli limited with enhanced Pauli fields compared to the bare weak coupling value of $\mu_0H_p=1.84\,T_\mathrm{c} \approx 0.5$\,T \cite{Clogston1962,Chandrasekhar1962}. We now proceed with the determination of the angle dependence of both the orbital limit $H_\mathrm{orb}$ and the Pauli limit $H_p$. In the clean-limit $H_\mathrm{orb}$ is given by the slope of the critical field $H_\mathrm{c2}(T)$ near $T_\mathrm{c}$ as $\mu_0H_\mathrm{orb}(T=0)=-0.73T_\mathrm{c}\left.\frac{dH_\mathrm{c2}}{dT}\right\vert_{T=T_\mathrm{c}}$\cite{WHH_clean_1966}.

From the data in Fig. \ref{fig:Figure_4} we extract the values of the $H_\mathrm{c2}$ slopes near $T_\mathrm{c}$ \cite{SM2021} and hence $H_\mathrm{orb}$ for all angles, which are given in Figure \ref{fig:Figure_2}b. We observe that $T_\mathrm{c}$ varies slightly even in zero field for different sample orientations, which is probably due to the remanent magnetic field in the superconducting magnet and slightly different shapes of the ac susceptibility curves at different angles near the onset of the transition where we define $T_\mathrm{c}$. The values of $T_\mathrm{c}$ are listed in \cite{SM2021}. Note that the uncertainty of the slope and hence $H_\mathrm{orb}$ at each angle is quite large (error bars in Figure \ref{fig:Figure_2}b). 

The angle dependence of $H_\mathrm{orb}$ behaves as expected for an anisotropic bulk superconductor \cite{Tinkham} according to
\begin{equation}
     H_\mathrm{c2}(\theta)=\frac{H_\mathrm{c2}^c}{\sqrt{\Gamma^2\sin{\theta}^2+\cos{\theta}^2}}
     \label{eq:Hc_in_angle}
\end{equation}
where the anisotropy parameter $\Gamma = H_\mathrm{c2}^c/H_\mathrm{c2}^{ab}$. We find $\Gamma =H_\mathrm{orb}^c/H_\mathrm{orb}^{ab}= 2.6$ and $\mu_0H_\mathrm{orb}^c=16.2$\,T as shown by the dashed orange line in Figure \ref{fig:Figure_2}b. For a fully gapped 3D superconductor in the Bardeen Cooper Schrieffer theory, this angle dependence reflects the anisotropy of the normal state since $\mu_0H_\mathrm{orb} = \Phi_0/2\pi\xi^2(T)$ and $\xi(0)=0.18\hbar v_\mathrm{F}/k_\mathrm{B}T_\mathrm{c}$. Using $v_{\mathrm{F}}=\hbar k_{\mathrm{F}}/m^\ast$ we find $H_\mathrm{orb}^c/H_\mathrm{orb}^{ab} = m_a^{\ast 2}/m_c^\ast m_a^\ast \Rightarrow m_a^\ast/m_c^\ast=2.6$. This is comparable to the weak anisotropy in the magnetization at 2\,K, where the in-plane value is a factor of 2 larger than the $c$-axis value \cite{Khim2020}. It will be interesting to see if band-structure calculations confirm this anisotropy in the future \cite{Hafner2021}.

As a next step, the Pauli paramagnetic limit of SC1 is investigated. Here we consider the following expression for the upper critical field of a spin-singlet superconductor with both orbital and Pauli limiting 

\begin{equation}
\ln(t)=\int\limits_0^{\infty}du \left\langle \frac{[1-F_{\theta}+F_{\theta}\cos(\frac{Hg_{\theta}u}{H_P t})]\exp(\frac{-Hu^2}{\sqrt{2}H_\mathrm{orb} t^2})-1}{\sinh u}\right\rangle  
\label{eq:2}
\end{equation}

where $g_{\theta}$ is a field-angle dependent $g$-factor, $t=T/T_\mathrm{c}$, and $F_{\theta}=1$ quantifies the pair-breaking due to Pauli limiting (as discussed later, this takes a different form for a pseudospin-triplet order parameter). $\langle ... \rangle$ indicates an angular average around the Fermi surface. Using the $H_\mathrm{orb}$ values from the fit of its angle dependence (in order to reduce the uncertainty at each angle), we find that $H_\mathrm{P}/g_{\theta}$ exhibits an anisotropy similar to $H_\mathrm{c2}(\theta)$, that has the angular-dependence given by Eq. \ref{eq:Hc_in_angle} with anisotropy parameter $\Gamma_g=2.8$. For more details about the fitting procedure please refer to \cite{SM2021}. Within error bars, there is no temperature dependence of the $H_\mathrm{c2}$ anisotropy for SC1.

Now let us turn to the critical field of SC2 with the aim to understand what is causing the strong suppression of $H_\mathrm{c2}$ when fields are turned away from the $c$ axis and the disappearance of SC2 at an angle of $35^\circ$. For $H\parallel c$, the temperature dependence of $H_\mathrm{c2}$ follows qualitatively - and even quantitatively, when a lower $T_\mathrm{c}$ is assumed - very well the expectations for a pure orbital limit \cite{Khim2020} without Pauli limiting effect \cite{SM2021}.

To first order, the orbital limit of SC2 should be similar to the one of SC1 given in Fig. \ref{fig:Figure_2}b\cite{footnote1}. Therefore, we can make the first statement that the angle dependence of the orbital limit is not strong enough to describe the steep decrease of the experimental critical field $H_\mathrm{c2}$ with angle. Hence, the decrease must be related to the Pauli limiting kicking in when the field is tilted towards the $ab$ plane. As discussed in the context of SC1, one place where the Pauli field introduces anisotropy is through the spin-orbit coupling renormalized $g$-factor $g_{\theta}$. However, this anisotropy is also too small to account for the steep decrease of the experimental critical field $H_\mathrm{c2}$ with angle since any renormalized $H_p$ for a spin-singlet state should have the same anisotropy as the one observed in SC1. Consequently, there must be another source of anisotropy due to Pauli limiting. Indeed, this is exactly the behavior expected for triplet superconductors with $\vec{d}$ vector in the plane (helical state): an absence of Pauli limiting for $H \parallel c$ (infinite $H_p^c$) and presence of Pauli limiting for $H \parallel ab$ which is isotropic in the plane. The latter is enhanced compared to the bare value \cite{Frigeri2004, Khim2020}, when Rashba  spin-orbit coupling is included. Here we consider such a helical triplet state subject to an orbital critical field, for which the critical field is given by the expression of Eq. \ref{eq:2} but now with $F_{\theta}=\vert\hat{d}\cdot\hat{h}_{\theta}\vert^2$, where $\hat{d}$ is $\vec{d}/\vert\vec{d}\vert$ and $\hat{h}_{\theta}$ is a unit vector that gives the direction of the Zeeman field projected onto to the pseudospin basis \cite{Khim2020}. As long as the orbital field energy ($g_{\theta}\mu_B H_\mathrm{orb}$) is smaller than the spin-orbit coupling energy, this is the expression for the odd-parity spin-singlet state in which the order parameter has opposite sign on the two inequivalent Ce-layers. In Fig. 2b, we show the calculated critical field for the state SC2. In the SM, we provide plots of the calculated $H-T$ phase diagrams for the different field angles. The agreement with experiment is excellent. Note that even the angle above which only one SC phase occurs is reproduced.

This leads us to the main conclusion of our paper: The angle dependence of the superconducting critical field is dominated by the huge anisotropy of the Pauli field of an odd parity state with d-vector in the plane, in agreement with the interpretation in \cite{Khim2020} and in previous models of locally non-centrosymmetric SC \cite{Yoshida2014}, although here the SC state is a pseudospin-triplet state with staggered dominant singlet intra-layer pairing. 

As a final point, we have also calculated $H^*$, the first-order transition between the SC1 and SC2 states. In the previous study for $H\parallel c$, the $H^*$ transition was found to be a result of a competition between the Pauli-limited SC1 (with Pauli field $H_\mathrm{P,1}$) and the Pauli-limit free SC2, so that $H^*\lessapprox H_\mathrm{P,1}$ \cite{Khim2020}. Here, we calculate the phase transition between SC1 and SC2 for all magnetic field directions from the free energies taking only the Pauli limiting effect into account (violet line in Fig. \ref{fig:Figure_2}b), with no additional parameters beyond those calculated from the fitting procedure. Neglecting the orbital limit leads to a slight overestimate of the critical field \cite{Schertenleib2021}, but the angular dependence is in perfect agreement with experiment \cite{SM2021}.

\section{Discussion}
We would like to emphasize that the simple model established in \cite{Khim2020} and extended here to intermediate angles can reproduce all experimental observations based on only 3 free parameters that cannot be measured experimentally: the critical temperature of the superconducting state SC2 $T_{c,2}$, the in-plane Pauli limit $H_\mathrm{P,1}$ and the strength of the Rashba spin-orbit coupling relative to the interlayer hopping $\tilde{\alpha}$. This model has the minimal number of bands in this crystal symmetry and naturally contains both even and odd parity superconducting states. Even though CeRh$_2$As$_2$ certainly contains multiple bands \cite{Hafner2021}, renormalized density functional theory calculations reveal the bulk of the density of states to be on symmetry-related Fermi surfaces near the zone boundary - which justifies the single-band approach used here \cite{Hafner2021,Cavanagh2021}.

The angle dependence of the orbital limit of SC1 corroborates a rather 3D Fermi surface in agreement with the strongly warped cylinders calculated with renormalised band-structure calculations \cite{Hafner2021}. Both the anisotropy of the effective mass given by the anisotropy of $H_\mathrm{orb}$ as well as the anisotropy of the g factor (and the magnetic susceptibility) are rather small. Furthermore, in quasi 2D systems such as Sr$_2$RuO$_4$\cite{Kittaka2009}, CeCoIn$_5$ \cite{Ikeda2001} or FeSe \cite{Farrar2020}, the critical field is usually larger for in-plane fields than for $c$-axis fields, because orbital motion perpendicular to the layers is hindered and almost no orbital limiting occurs with large $H_\mathrm{orb}$ for in-plane fields. In the 2D limit a cusp is expected for $H_\mathrm{orb}$ for fields close to the $a$ axis as observed for example in FeSe, K$_2$Cr$_3$As$_3$ or in superlattices of CeCoIn$_5$/YbCoIn$_5$ \cite{Tinkham,Farrar2020,Zuo2017,Goh2012}, but not observed here. A previous theoretical study found that going from a quasi two dimensional Fermi surface to a three dimensional Fermi surface reduces the anisotropy of SC1 due to Rashba spin-orbit coupling, but shouldn't affect qualitatively the anisotropy of SC2 for the suggested scenario here \cite{Skurativska2021}. However, they investigated a Fermi surface at the Brillouin zone center and the situation may change for a Fermi surface at the zone boundary where -- due to the non-symmorphic structure and symmetry-imposed degeneracies -- large values of the Rashba strength over interlayer coupling are expected \cite{Cavanagh2021}.

In ferromagnetic superconductors, which are the most prominent candidates of spin-triplet superconductivity, the critical fields are highly anisotropic \cite{Aoki2009, Aoki2019}, and the strong enhancements observed along certain directions or in the field-reentrant phases are related with strong Ising-type spin fluctuations and quantum criticality influencing the orbital limit \cite{Aoki2019,Tada2011}.  Here we find that this is not the case for CeRh$_2$As$_2$, where the anisotropy is accounted for by a pseudospin $\vec{d}$ that is oriented in the basal plane. An interesting open issue is why the orbital critical field and the Sommerfeld coefficient are so large for CeRh$_2$As$_2$. In the non-centrosymmetric superconductors  CePt$_3$Si and CeRhSi$_3$ the origin of this was proposed to be antiferromagnetic quantum critical fluctuations that enhance the electron effective mass \cite{Kimura2005,Kimura2007,Sugitani2006,Bauer2012,Tada2011}. This needs to be investigated further in CeRh$_2$As$_2$, where quadrupolar as well as antiferromagnetic degrees of freedom are suggested to play a role \cite{Hafner2021, Kibune2021,Kitagawa2022}. CeIrSi$_3$, CeRhSi$_3$ and CeCoGe$_3$ show a similar anisotropy of the superconducting state to CeRh$_2$As$_2$ with absence of Pauli limit for $H\parallel c$ and strong but enhanced Pauli limit for $H\parallel ab$ \cite{Kimura2007,Settai2008,Measson2009}, but a full determination of the angle dependence hasn't been possible yet because the large anisotropy appears only under pressure, preventing an easy angle-dependent measurement.

At this point, we would like to discuss the possibility that the transition between the two superconducting states originates from the suppression of the $T_0$ order~\cite{Hafner2021,Khanenko2022} or of the AFM order~\cite{Kibune2021} by a field along the $c$ axis. In the former scenario, the suppression of the $T_0$ order is considered to be responsible for the $H^*$ transition line and the superconducting order parameter remains the same in the SC1 and SC2 states, i.e. below and above $H^*$. Our results indicate that whenever the superconducting state coexists with the $T_0$ order, it is Pauli limited, but when $T_0$ is suppressed for fields larger than $\mu_0H^* \approx 4$\,T, the superconducting state is not anymore, or much less, Pauli limited. Knowing that the $T_0$ order probably causes a nesting, i.e. partial gapping of the Fermi surface, it seems natural that it might affect the superconducting state. Accordingly, in order to understand the observed anisotropy of the superconducting state in this paper, the $T_0$ order would at least have to suppress the spin-orbit coupling, for example by gapping out parts of the Fermi surface with large spin-orbit coupling. However, this seems difficult to explain microscopically, as it would imply that the bare spin-orbit coupling is significantly larger than the already substantial spin-orbit coupling used to fit the critical field data in Fig.~\ref{fig:Figure_2}. In the latter scenario, the AFM order is considered to be responsible for the $H^*$ line. The behavior of the AFM state with magnetic field and angle is not known yet, besides a single nuclear magnetic resonance (NMR) measurement showing a line-broadening starting between 0.2 and 0.3\,K at 1.4\,T along the $c$ axis (Supplementary material of \cite{Kibune2021}). The AFM order is expected to be suppressed by the magnetic field and the angle dependence of this suppression might be similar to the $H^*$ line. However, with a transition temperature of $T_\mathrm{N} \approx 0.25$\,K in zero field, it seems very unlikely that the suppression of $T_\mathrm{N}$ would be responsible for the transition line at $\mu_0H^* = 4$\,T. In fact, this would imply $T_\mathrm{N}$ to be almost constant up to 4\,T followed by a first order phase transition. While such behavior is unexpected, it has been observed e.g., in systems which are near a ferromagnetic QCP and show field induced tricritical points, like Yb(Rh$_{1-x}$Co$_x$)$_2$Si$_2$ with $x=0.18$~\cite{Hamann2019}: In this material the zero-field AFM phase decreases only very little with field along the $c$ axis, and then a first-order transition to a polarized state occurs. However, the underlying physics of this system is different, since it is antiferromagnetic but very close to a ferromagnetic state with an extremely large magnetocrystalline anisotropy of $\approx 10$. In CeRh$_2$As$_2$ the susceptibility is too small to be close to ferromagnetism and the magnetocrystalline anisotropy is only $\lessapprox 2$. So it is unlikely that similar physics is at play in CeRh$_2$As$_2$.

Although the angle-dependence cannot completely rule out these scenarios, a transition between two superconducting states seems more natural, especially given how well it fits the data.

\section{Conclusion} 
The excellent agreement of the model with the experimental results strengthens the interpretation that the superconducting state changes from even to odd parity at $H^*$ and that the strong anisotropy of the critical field is rooted in the Pauli limiting effect of a helical pseudospin $d$-vector. Since other orders are not included in this model and are not needed to obtain the good agreement, their influence on the superconducting phase diagram and especially on $H^*$ appears to be small. However, this point can only be resolved and other scenarios ruled out when more microscopic information on the orders and on the pairing mechanism will be available in the future, so that their interplay can be investigated and understood.

\section{Acknowledgements}
We thank Konstantin Semeniuk, Aline Ramirez, David Moeckli, and Mark Fischer for stimulating discussions.
JL and EH acknowledge support from the Max-Planck society for funding of the Max Planck research group "Physics of unconventional metals and superconductors". Torque measurements were carried out at the European High Magnetic Field lab in Grenoble.
CG and EH are also supported by the joint Agence National de Recherche (ANR) and DFG program Fermi-NESt through grant GE602/4-1.
DCC and PMRB were supported by the Marsden Fund Council from Government funding, managed by Royal Society Te Ap\={a}rangi. 
DFA was supported by the US Department of Energy, Office of Basic Energy Sciences, Division of Materials Sciences and Engineering under Award DE-SC0021971.
\section{Supplementary material}
\subsection{Methods}
\subsubsection{Samples}
Single crystals of CeRh$_2$As$_2$ were grown in Bi flux as described elsewhere \cite{Khim2020}. For ac-susceptibility and torque, we used the same batch as in \cite{Khim2020}. For the specific heat, a different batch with almost identical $T_c$ and $T_0$ was used.

\subsubsection{Ac-susceptibility}
The magnetic ac susceptibility $\chi$ was measured using a homemade set of compensated pick-up coils as described in \cite{Khim2020}. A superconducting modulation coil built in the main magnet produced an excitation field of 175\,$\mu$T at 5\,Hz. The output signal of the pick-up coils was amplified using a low temperature transformer (LTT-m from CMR) with a winding ratio 1:100, a low noise amplifier SR560 from Stanford Research Systems and finally measured in a digital lock-in setup using a 24 bits PXI2-4492 as data acquisition system \cite{Khim2020}. The ac susceptibility measurements were performed in a MX400 Oxford dilution refrigerator down to 45\,mK and up to 15\,T. For the angle dependence of $\chi$ we used a Swedish rotator installed in the dilution refrigerator and both the pick-up coil and the sample rotate together. The angle between the external magnetic field and the sample is defined with respect to the tetragonal $c$ axis of a small single crystal of CeRh$_2$As$_2$ of a volume of $\sim  500\times 500 \times 500\,\mu$m$^3$. The data from temperature sweeps were normalized to their respective value in the normal state at 0.5\,K for all the magnetic fields applied. For the field sweeps, the absolute value of the signal at different temperatures is given, normalized to the curve measured at 0.45\,K.

\begin{figure}[h]
	\centering
	\includegraphics[width=0.9\linewidth]{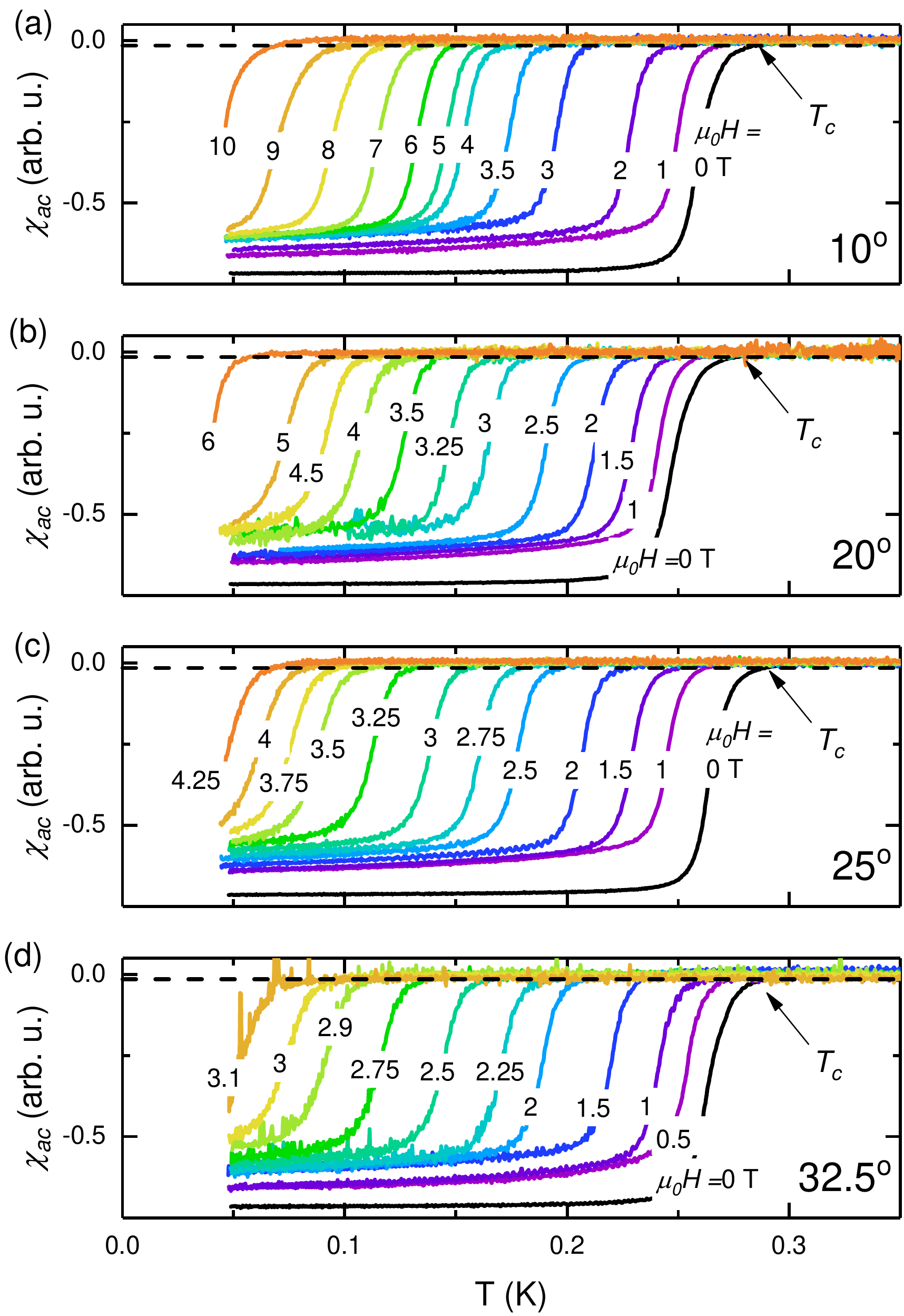}
	\caption{{Temperature dependence of ac magnetic susceptibility at differents magnetics field.(a)(b)(c)(d) are the data for $10^\circ$, $20^\circ$, $25^\circ$ and $32.5^\circ$ respectively. The dashed line shows where $T_c$ is defined.}}
\label{fig:Figure_AC_SM}
\end{figure}

The Figure \ref{fig:Figure_AC_SM} shows the temperature dependence of $\chi$ with an external magnetic field $\mu_0 H$ applied at different angles. The dashed lines show the value of $\chi$ where $T_c$ is defined as the onset of the diamagnetic drop.

\subsubsection{Torque}
The magnetic torque $\vec{\tau} = \vec{M} \times \vec{B}$ was measured using a 50\,$\mathrm{\mu}$m thin CuBe cantilever via capacitive readout in a top-loading dilution refrigerator at base temperature in the M9 magnet at the Laboratoire National des Champs Magn\'etiques Intenses. A tiny piece of the resistivity sample from \cite{Khim2020, Hafner2021} was used. The cantilever plane was parallel to the field for $H\parallel [110]$ and then rotated to the perpendicular orientation where $H\parallel c$ axis. The field was swept always in the same direction out of the superconducting state and back to zero. For data analysis, we subtracted the zero-field capacitance at each angle and then plot the capacitance change as a function of field, which is proportional to the torque. We observe that in the normal state, the angle and field dependence follows rather well the expected $\sin(2\theta)$ and $H^2$ dependencies, respectively, where $\theta$ is again the angle measured from the c-axis. A scaling of the data with $\sin(2\theta)$ makes the torque at different angles comparable and dividing by $H^2$ makes deviations from the normal state behavior more visible, especially when superconductivity sets in. The true torque can in principle be obtained by multiplying with a factor including the spring constant of the cantilever and the position of the sample. Here however, we leave the arbitrary units since we focus on the anomalies at phase transitions.

\subsubsection{Specific heat} 

The measurements were carried out with a quasi adiabatic heat pulse technique as described in \cite{Wilhelm2004}. Compared to the data presented in \cite{Khim2020} where specific heat was measured on a sample with 5\,mg mass, the sample here weighed 19\,mg. Although samples were from different growth batches, the data in zero field lie exactly on top of each other. The sample was fixed to the platform with wedges of the desired angles. The uncertainty of angle in this method is estimated to $\pm 2.5^\circ$.

\begin{figure}[h]
	\centering
	\includegraphics[width=0.9\linewidth]{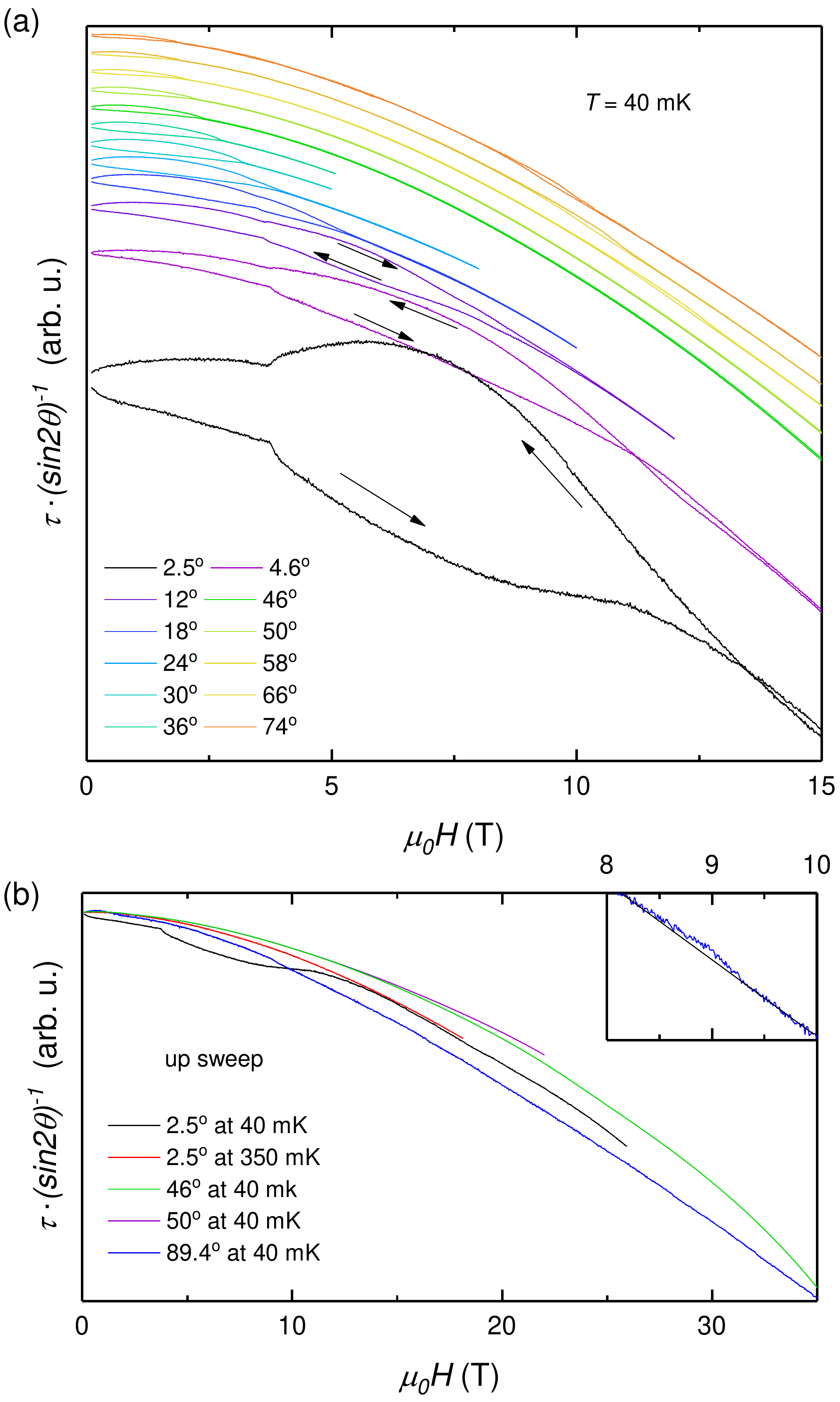}
	\caption{Magnetic field dependence of $\tau/\sin(2\theta)$ for angles as indicated.(a) a strong increase of the hysteresis loop is observed for $2.5^\circ$. (b) no other phase transitions than $H_q$ and the superconducting state are observed up to 36\,T for $H\parallel ab$ and up to 26\,T for $H\parallel c$. The inset shows the $H_q$ transition at 9.2\,T for an angle of $0.6^\circ$ away from $H\parallel ab$. }
\label{fig:Figure_Torque_SM}
\end{figure}

\subsection{Change of hysteresis loop in the magnetic torque for angles close to the $c$ axis}

\begin{figure}[t]
	\centering
	\includegraphics[width=0.9\linewidth]{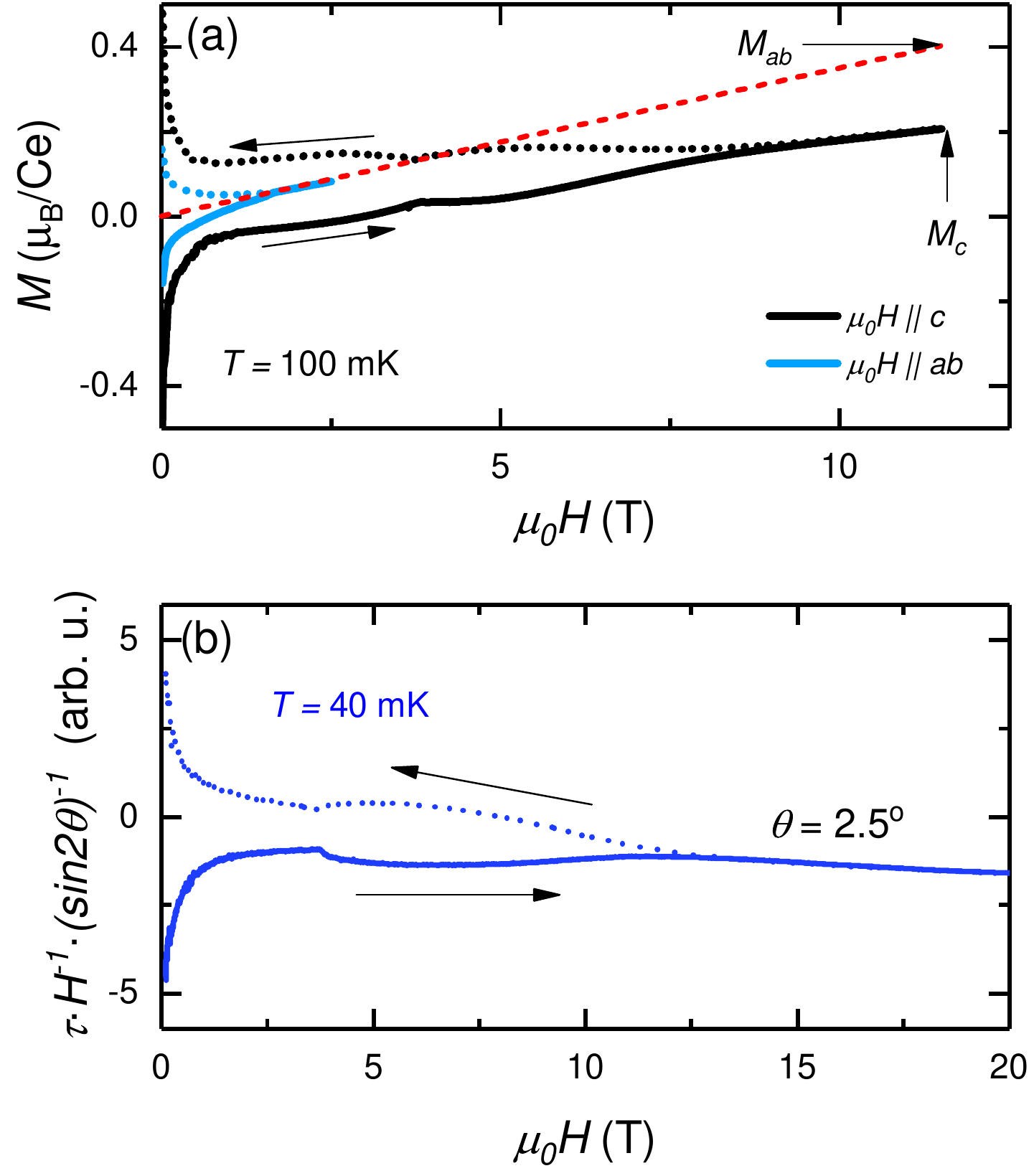}
	\caption{(a) DC-magnetization along c and ab at 100 mK. The red line is the extrapolation of the magnetization to 12 T. (b) is the torque divided by the field, note that $\tau/H\propto M$.}
\label{fig:Figure_M}
\end{figure}
The raw data of magnetic torque for selected magnetic field angles is shown in Fig. \ref{fig:Figure_Torque_SM}. In the superconducting state, a strong hysteresis loop is observed for all angles as already observed in the magnetization (shown in Fig. \ref{fig:Figure_M}a). We observe a change in the hysteresis loop which surprisingly changes sign as a function of field angle. For $2.5^\circ$ and 4.6$^\circ$ the up-sweep torque in the superconducting state joins the normal state behavior from the bottom. In contrast, for all larger angles, the up-sweeps approach the normal state behavior from the top (Fig. \ref{fig:Figure_Torque_SM}a). 
For fields very close to the c-axis, the ratio of torque and magnetic field, $\tau/H$, is expected to follow the magnetization. Our experimental result nicely follows this expectation as shown in Fig. \ref{fig:Figure_M} where we compare $M(H)$ for field along $c$ (Fig. \ref{fig:Figure_M}a) with $\tau/H$ at 2.5$^\circ$ (Fig. \ref{fig:Figure_M}b) \cite{Khim2020}. 
The amplitude of the torque is proportional to the difference of the susceptibilities $\chi_a-\chi_c$. Since the magnetization is linear in field for both directions in the normal state with $M_{ab} \approx 2 M_c$ as depicted in Fig. \ref{fig:Figure_M}a, the normal state susceptibility has the same anisotropy as the magnetization. In the superconducting state at low angles, this difference is larger than in the normal state, implying that the hysteresis loop is larger in $M_c$ than in $M_a$. At larger angles, it is the inverse, meaning that the hysteresis loop is larger in $M_a$ than in $M_c$. This change causes the difference in the hysteresis loop. Since the hysteresis loop is caused by vortex physics, this is a sign of a change of the anisotropy of the vortex state with magnetic field direction with a strong enhancement of the hysteresis loop in $M_c$ when the magnetic field is close to the c-axis.

In Fig. \ref{fig:Figure_Torque_SM}a, the transition at $H_{cr}$ is also visible.

Fig. \ref{fig:Figure_Torque_SM}b shows field sweeps up to the highest measured fields at some angles. At 46$^\circ$ and 50$^\circ$  no other transition than $H_{c2}$ is observed.

\subsection{Fits}
\begin{figure}[t]
	\centering
	\includegraphics[width=0.9\linewidth]{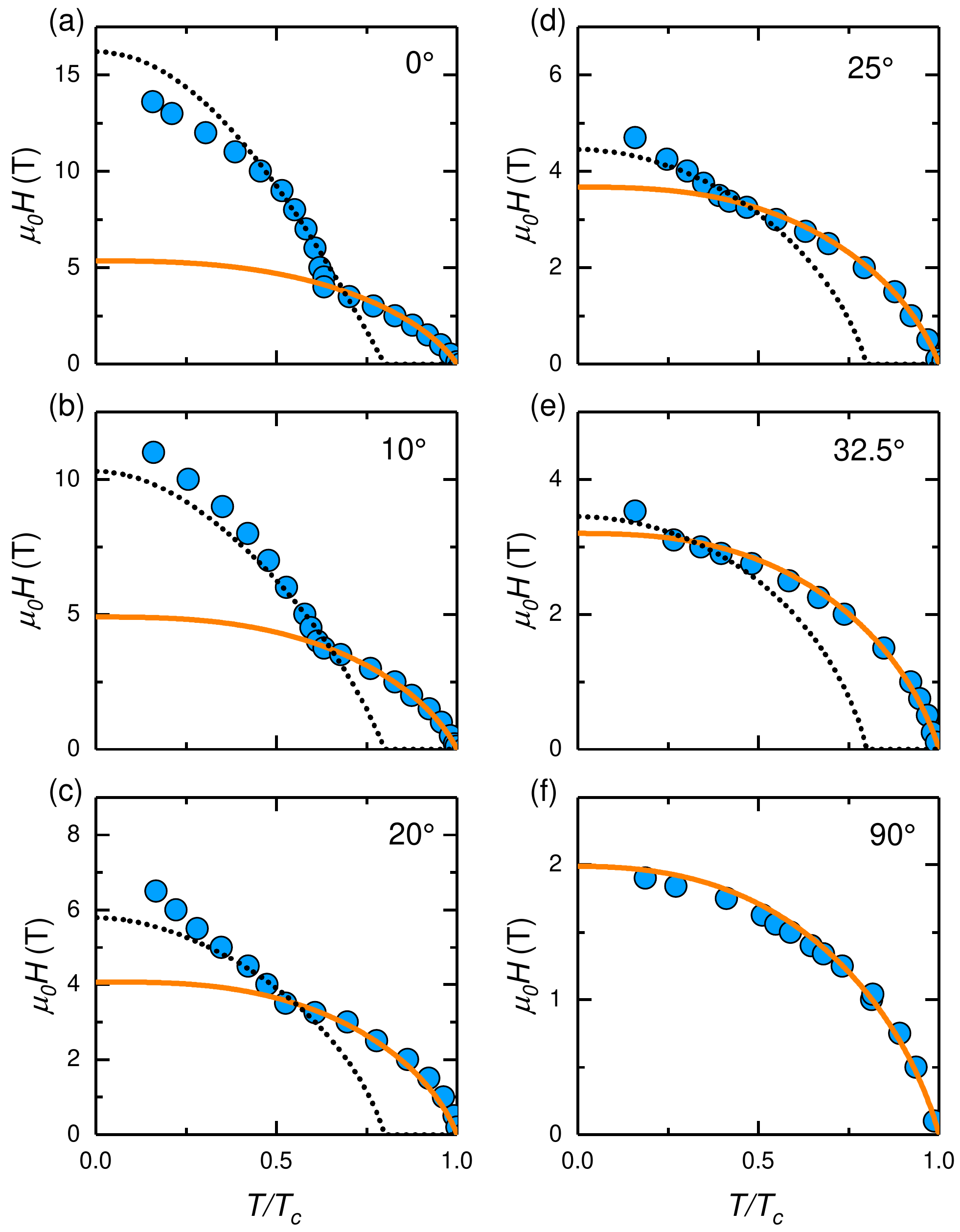}
	\caption{The blue circles are the $T_c$ obtained form the temperature dependence of the ac-susceptibility (Figure \ref{fig:Figure_AC_SM}) at different magnetic fields for angles of 10, 20, 25 and 32.5. For 0 and 90 the data were taken from \cite{Khim2020}. The orange full lines are the fits for SC1, whereas the black doted lines are the expected behavior of SC2.}
\label{fig:Fits}
\end{figure}

\begin{figure}[t]
	\centering
	\includegraphics[width=0.9\linewidth]{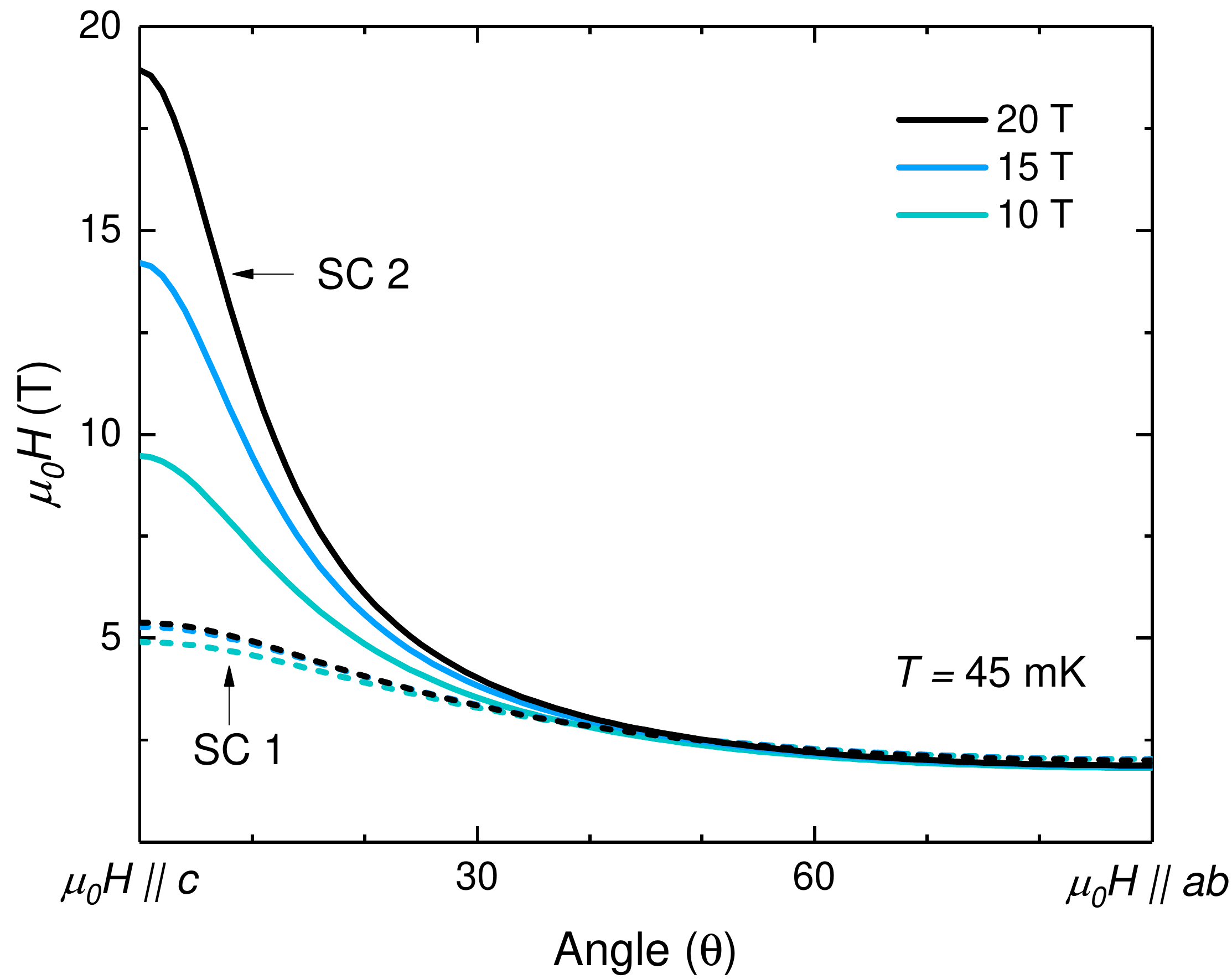}
	\caption{Angle dependence of the upper critical field at 45 mK for SC1 (dashed lines) and SC2 (solid line) using an angle-independent $H_{orb}$ of 20 T, 15 T and 10 T.}
\label{fig:Constant Horb}
\end{figure} 
To perform the theoretical fit to the experimental data of both the angle dependence at low temperature and the temperature dependence at fixed magnetic-field angle, we solve the linearized gap equation in the presence of both Pauli and orbital limiting effects,

\begin{equation}
\ln(t)=\int\limits_0^{\infty}du \langle \frac{[1-F_{\theta}+F_{\theta}\cos(\frac{Hg_{\theta}u}{H_P t})]\exp(\frac{-Hu^2}{\sqrt{2}H_{orb} t^2})-1}{\sinh u}\rangle  
\end{equation}
as discussed in the main text, where $\langle\ldots\rangle$ denotes the average over the Fermi surface, $H_P$ is the Pauli limiting field for in-plane fields, $H_{orb}$ the orbital limiting field, $t=T/T_c$, $g_{\theta}$ is the (angle-dependent) effective $g$-factor, and $F_{\theta}$ quantifies the pair-breaking effects of the Pauli field. For an even-parity superconductor, $F_{\theta}=1$ as the field will always be completely pair-breaking. On the other hand, for odd-parity states, $F_{\theta} =|\hat{d}\cdot\hat{h}_{\theta}|^2$, where $\hat{d}$ is the normalized vector describing the pseudospin structure of the superconducting gap, and $\hat{h}_{\theta}$ is a unit vector that gives the direction of the Zeeman field in the pseudospin basis. Therefore $0\leq F_{\theta}\leq 1$. The anisotropy of the Pauli limiting comes from both the intrinsic anisotropy of the $g$-factor as measured by the susceptibility \cite{Khim2020}, and from the spin-orbit coupling which introduces an additional anisotropy in the intraband component of the effective $g$-factor, relevant for Pauli limiting, as well as in the pair-breaking parameter $F_{\theta}$. The forms of both $g_{\theta}$ and $F_{\theta}$ are determined by the spin-orbit coupling,

\begin{align}
g_{\theta} &= \sqrt{\frac{\left(g_{c,0}^2\cos^2\theta+g_{ab,0}^2\sin^2\theta\right)+g_{ab,0}^2\tilde{\alpha}^2\sin^2\theta\sin^2\phi}{1+\tilde{\alpha}^2}}\nonumber\\
F_{\theta} &= \frac{g_{ab,0}^2\left(1+\tilde{\alpha}^2\right)\sin^2\theta\sin^2\phi}{\left(g_{c,0}^2\cos^2\theta+g_{ab,0}^2\sin^2\theta\right)+g_{ab,0}^2\tilde{\alpha}^2\sin^2\theta\sin^2\phi},
\end{align}
where $\tilde{\alpha}$ is the strength of the Rashba spin-orbit coupling relative to the interlayer hopping \cite{Khim2020}, $\theta$ is the angle from the $c$-axis at which the field is applied, $\phi$ is the angle around the Fermi surface which for simplicity we assume consists of a single circular sheet located at the centre of the Brillouin zone. $g_{ab,0}$ and $g_{c,0}$ are the values of the $g$-factor for in-plane and $c$-axis fields, respectively.  The anisotropy in $F_{\theta}$ plays a much more important role than that of $g_{\theta}$. In particular, for fields in plane, $F_{\theta}$ implies that there is Pauli limiting. However, for all other field orientations, $F_{\theta}$ implies a divergence in the Pauli field in the zero temperature limit. As temperature approaches zero, this divergence is strongly angle dependent: it is weak for fields close to the basal plane and strong for fields close to the $c$-axis. This divergence is cut-off by the orbital field in our theory, but the underlying anisotropy of this divergence still manifests itself, and is the main source of the anisotropy in the SC2 phase. This is reflected in Fig. \ref{fig:Constant Horb} where $H_{orb}$ is taken as a constant value and the critical field of SC2 is nevertheless highly anisotropic.

To determine the best fit to the data, we first consider the even-parity SC1, for which we set $F_{\theta}=1$. Using the orbital fields determined from the experimental data at each angle (as described in the main text), we vary the in-plane Pauli limiting field $H_P$ and spin-orbit coupling strength $\tilde{\alpha}$ for all angles simultaneously. The best fit to the SC1 data was found with $H_{P,1}=2.3$\,T and $\tilde{\alpha}=2.7$. These fits are given as orange lines in Fig. \ref{fig:Figure_AC_SM}. Differences in values for fit parameters compared to \cite{Khim2020} are due to the fact that we fit data from ac-susceptibility here (and not specific heat).

Turning to the odd-parity SC2, we must additionally determine the critical temperature $T_{c,2}$, relative to the critical temperature for SC1 $T_{c,1}$. As such, we fit the SC2 data by varying the ratio $T_{c,2}/T_{c,1}$, with the spin-orbit coupling unchanged and the Pauli limiting field reduced from the SC1 value by the scaling $H_{P,2}=H_{P,1}(T_{c,2}/T_{c,1})$, where $H_{P,1}=2.3$\,T is the Pauli field for SC1, and $H_{P,2}$ the Pauli field for SC2. In order to keep the amount of parameters as low as possible, we assume that the orbital fields for SC2 take the same value as those for SC1. After determining the optimal $T_{c,2}=0.80T_{c,1}$, we finally allow the orbital fields to vary within the experimental margin of error, to further improve the fits of the temperature dependence. The resulting fits are shown in Fig. \ref{fig:Fits} as black dashed lines, and are qualitatively robust against small variations in the fitting  parameters.

To calculate $H^*$, the first order transition between SC1 and SC2, we calculated the free energy in the absence of orbital limiting, via the method described in Ref.  \cite{Cavanagh2021} using the model parameters $H_{P1}$, $T_{c2}/T_{c1}$ and $\tilde{\alpha}$ found from the procedure described above, with no additional fitting parameters. We include the contribution to the SC2 free energy due to the Pauli limiting effect when $\theta \neq 0$, accounted for by a modification of the superconducting quasiparticle energy defined by $F_{\theta}$, 
\begin{equation}
E_{\sigma}=\sqrt{\left(g_\theta \frac{H}{H_P}\right)^2 + \xi^2 +\bar{\Delta}^2 +2\sigma\sqrt{\left(g_\theta \frac{H}{H_P}\xi\right)^2+F_\theta\bar{\Delta}^2}},
\end{equation}
where $\xi$ is the normal state energy dispersion and $\bar{\Delta}$ is the magnitude of the gap projected onto the Fermi surface. The calculated transition line $H^*$ is shown in Fig.2b, where it clearly agrees excellently with the angle dependence of the transition.

\begin{table}[h]
  \begin{center}
\begin{tabular}{c|cccccc} 
\hline
\hline
Parameter &0$^\circ$&10$^\circ$&20$^\circ$&25$^\circ$&32.5$^\circ$&90$^\circ$\\
\hline
$T_c$ (K)&0.286&0.280&0.275&0.284&0.282&0.293\\
$-\left.dH/dT\right|_{T_c}$ (T/K) &80&77&71&57&50&27\\
$H_{orb}$ (T)&16.7&15.7&14.3&11.8&10.3&5.8\\
$H_{orb}$ (T), fit &16.2&15.7&14.7&11.8&10.3&5.8\\
\hline
\hline
    \end{tabular}
      \caption{Experimental values of $T_c$,$-dH/dT|_{T_c}$ and $H_{orb}$, and $H_{orb}$ fit parameters. From the fitting, we find $\tilde{\alpha}=2.7$, $H_{P,1}=2.3T$ for SC1, and $H_{P,2}=1.84T$ for SC2, with $T_{c,2}/T_{c,1}=0.80$ the ratio between  the critical temperatures of SC2 and SC1.}\label{Tab1}
  \end{center}
\end{table}


%

\end{document}